# A Study of Implanted and Wearable Body Sensor Networks

Sana Ullah[1], Henry Higgin[2], M. Arif Siddiqui[1], and Kyung Sup Kwak[1]

[1] Graduate School of Telecommunication Engineering
253 Yonghyun-Dong, Nam-Gu, 402-751, Inha University Incheon South Korea
sanajcs@hotmail.com, arif.sid@hotmail.com, kskwak@inha.ac.kr
[2] Zarlink Semiconductor Company
Caldicot NP26 5YW United Kingdom
henry.higgins@zarlink.com

**Abstract.** Recent advances in intelligent sensors, microelectronics and integrated circuit, system-on-chip design and low power wireless communication introduced the development of miniaturised and autonomous sensor nodes. These tiny sensor nodes can be deployed to develop a proactive Body Sensor Network (BSN). The rapid advancement in ultra low-power RF (radio frequency) technology enables invasive and non-invasive devices to communicate with a remote station. This communication revolutionizes healthcare system by enabling long term health monitoring of a patient and providing real time feedback to the medical experts. In this paper, we present In-body and On-body communication networks with a special focus on the methodologies of wireless communication between implanted medical devices with external monitoring equipment and recent technological growth in both areas. We also discuss about open issues and challenges in a BSN.

## 1 Introduction

The leading cause of death in US is heart disease, i.e. about 652,486 and 150,074 people died due to cardiovascular and cerebrovascular diseases [1]. In South Korea, 17% people die due to cerebrovascular diseases [2]. The health care expenditure in US is expected to reach 2.9 trillion by 2009 and 4 trillion by 2015, or 20% of Gross Domestic Product (GDP) [3]. Cardiovascular disease is the leading cause of death and it accounts for approximately 30% of all deaths worldwide [4]. In UK, it is 39% of all deaths [5]. In Europe, 90% of people die due to arrhythmogenic event [6]. Irregular heart beat causes such deaths and can be monitored before heart attack. Holter monitors are used to collect cardio rhythm disturbances for offline processing without real time feedback. Transient abnormalities are sometimes hard to capture. For instance, many cardiac diseases are associated with episodic rather than continuous abnormalities such as transient surges in blood pressure, paroxysmal arrhythmias or induced episodes of myocardial ischemia and their time cannot be predicted [6]. The accurate prediction of these episodes improves the quality of life.

Body Sensor Network (BSN) is a key technology to prevent the occurrence of myocardial infarction, monitoring episodic events or any other abnormal condition and can be used for long term monitoring of patients. The term BSN is first coined by Prof Guang-Zhong Yang of imperial college London. A BSN consists of miniaturised, low power and noninvasive or invasive wireless biosensors, which are seamlessly placed on or implanted in human body in order to provide an adaptable and smart healthcare system. This seamless integration of small and intelligent wireless sensors is used to monitor the patient's vital signs, provide real time feedback and can be a part of diagnostic procedure, maintenance of chronic condition, supervised recovery from a surgical procedure and to monitor effect of drugs therapy [7]. Each tiny biosensor is capable of processing its own task and communicates with a network coordinator or PDA. The network coordinator sends patient information to a remote server for further analysis. Episodic examination of a patient captures a snapshot of recovery process and skips other potential complications [8]. A BSN focuses on early detection of life threatening abnormalities and maintenance of chronic condition [9]. Long term monitoring of patient activities under natural physiological states improves quality of life by allowing patients to engage in normal daily life activities, rather than staying at home or hospital [10]. Moreover, implants for therapeutic and diagnostic purposes are also becoming more common. They can be used to restore control over paralyzed limbs, enable bladder and bowel muscle control, maintain regular heart rhythm, and many other functions. These implants significantly improve the quality of life of many patients. Though BSN research is in inception, but a number of on going research has enabled the innovation of several prototypes for unobtrusive pervasive healthcare system. However, no standard exists for a BSN due to considerable number of issues and challenges such as interoperability, privacy and security, low power communication, biosensor design, baseline power consumption, communication link between the implanted device and external monitoring and control equipment. The scope of a BSN spans around three domains: Off-body communication, On-body communication and In-body communication. Off-body communication is the communication from the base station to the transceiver on human side. On-body communication is the communication within on-body networks and wearable system. In-body communication is the communication between invasive or implantable devices with a base station.

The rest of the paper is divided into four categories. Section 2 contains a detailed discussion on invasive or in-body communication with a special focus on the methodologies of wireless communication between implanted medical devices with external monitoring equipment. This includes discussion on inductive coupling, in-body RF communication and antenna design. Section 3 presents a brief discussion on non-invasive or on-body communication and the recent technological growth in this area. Section 4 contains discussion on multi-agent technology for a BSN. Section 5 focuses on open issues and challenges in a BSN. Finally, we present conclusion to our work.

## 2 In-Body Communication

Advancement in implant technology and RF communication has enabled the communication of invasive or implanted device with a remote base station and can monitor every aspect of a patient. These new implant technologies require a communication link between the in-body device and external monitoring and control equipment. Zarlink semiconductor has introduced the world's first wireless chip, which supports a very high data rate RF link for communication with an implantable device [11]. The ZL70101 ultra-low power transceiver chip satisfies the power and size requirements for implanted communication systems and operates in 402-405 MHz Medical Implantable Communications Service (MICS) band [12].

There are several ways to communicate with a human body implant, including methods that use electromagnetic induction (similar to radio frequency identification, or RFID) and the others that use radio frequency (RF). Both are wireless and their use will depend on applications.

### 2.1 Inductive Coupling

Several applications still use electromagnetic coupling to provide a communication link to implanted devices, with an external coil held very close to the patient that couples to a coil implanted just below the skin surface. The implant is powered by the coupled magnetic field and requires no battery for communication. As well as providing power, this alternating field is also be used to transfer data into the implant. Data is transferred from the implanted device by altering the impedance of the implanted loop that is detected by the external coil and electronics. This type of communication is commonly used to identify animals that have been injected with an electronic tag. Electromagnetic induction is used when continuous, long-term communication is required.

The base band for electromagnetic communication is typically 13.56 MHz or 28 MHz, with other frequencies also available. Its use is subject to regulation for maximum Specific Absorption Rate (SAR). Inductive coupling achieves the best power transfer when using large transmit and receive coils, meaning it's impractical when space is an issue or devices are implanted deep within the patient. This technique does not support a very high data rate and cannot initiate a communication session from inside of the body. Please read Finkenzeller et.al [13] for further details.

### 2.2 In-Body RF-Communication

Compared with inductive coupling, RF communication dramatically increases bandwidth and enables a two-way data link to be established. The Medical Implant Communication Service (MICS) band of 403 MHz to 405 MHz is gaining worldwide acceptance for in-body use [14]. This band has a power limit of 25 µW in air and is split into 300 kHz wide channels. The human body is a medium that poses numerous wireless transmission challenges. The body is composed of varied components that are not predictable and will change as the patient ages, gains or loses weight, or even changes posture. There are formulas for designing free-air communications but it's very difficult to calculate performance for an in-body communication system. To compound the design challenge, the location of the implanted device is also variable. During surgery the implant is placed in the best position to perform its primary function, with little consideration for wireless performance.

Typical dielectric constant ($\varepsilon_r$), conductivity ($\rho$) and characteristic impedance $Z_0(\Omega)$ properties of muscle and fat are shown in Table 1.

**Table 1.** Body Electrical Properties

| Frequency | Muscle | | | Fat | | |
|---|---|---|---|---|---|---|
| | ($\varepsilon_r$) | $\rho$ (S.m$^{-1}$) | $Z_0(\Omega)$ | ($\varepsilon_r$) | $\rho$ | $Z_0(\Omega)$ |
| 100 | 66.2 | 0.73 | 31.6 | 12.7 | 0.07 | 92.4 |
| 400 | 58 | 0.82 | 43.7 | 11.6 | 0.08 | 108 |
| 900 | 56 | 0.97 | 48.2 | 11.3 | 0.11 | 111 |

The dielectric constant has an effect on the wavelength of a signal. In air the wavelength can be found from Equation 1 where $\varepsilon_r = 1$. However in a different medium the wavelength is reduced as in Equation 2.

$$\lambda = 300\frac{10^6}{f} \tag{1}$$

where $\lambda$ is the wavelength in air in meters and $f$ is frequency in Hz.

$$\lambda_{medium} = \frac{\lambda}{\sqrt{\varepsilon_r}} \tag{2}$$

where $\lambda_{medium}$ is the wavelength in medium.

At 403 MHz the wavelength in air is 744 mm, but in muscle with $\varepsilon_r = 50$ the $\lambda_{medium} = 105$ mm. This is of considerable help in designing implanted antennas where physical size is an important consideration. The conductivity of muscle is 0.82Sm$^{-1}$ – this is more than air, which is almost zero. The effect of this is similar to surrounding the implant with seawater that will attenuate the signal as it passes through. This results in reduced penetration. The characteristic impedance $(Z_o)$ is relevant when it changes, such as at the fat-muscle boundary. This will cause part of the signal to be reflected by a term known as reflection co-coefficient Γ, found from Equation 3.

$$\Gamma = \frac{Z_o - Z_r}{Z_o + Z_r} \tag{3}$$

where $Z_o$ is the impedance of free space (377 Ω), and $Z_r$ is the impedance of medium in Ω. This results in a signal being reflected of magnitude Γ of incident signal power. So for muscle-fat boundary Γ = 80% of the signal is reflected. As an implant does not have an earth (ground), the case or other wires will also radiate. This means that signals will be radiated from the antenna and other structures associated with the implant. More details are available in Yang et.al [15].

## 2.3 Antenna Design

An in-body antenna needs to tuneable, using an intelligent transceiver and routine. This will enable the antenna coupling circuit to be optimised and the best signal strength obtained. Often size constraints dictate the choice of a non-resonant antenna. A non-resonant antenna will have lower gain and therefore be less sensitive on receiving and radiate less of the power generated by the transmitter. This makes design of the antenna coupling circuit even more important.

A patch antenna can be used when the implant is flat and there is no room to deploy a short wire. Patch antennas comprise a flat substrate coated on both sides with conductor. The substrate is typically alumina or a similar body-compatible material, with platinum or platinum/iridium coating both surfaces. The upper surface is the active face and is connected to the transceiver. The connection to the transceiver needs to pass through the case where the hermetic seal is maintained, requiring a feed-through. The feed-through must have no filter capacitors present; these are common on other devices. A patch antenna will be electrically larger than its physical size because it will be immersed in a high $\varepsilon_r$ medium. It can be made to appear even larger electrically if the substrate is of high $\varepsilon_r$ . The off-resonance antennas have low radiation resistance, typically in the order of a few Ohms for a patch. A loop antenna is an option where it can be deployed attached to the implant case. The loop antenna operates mostly with the magnetic field, whereas the patch operates mostly with the electric field. The loop antenna delivers comparable performance to that of a dipole, but with a considerably smaller size. Also the magnetic permeability of muscle or fat is very similar to that of air, unlike the dielectric constant that varies considerably. This property enables an antenna to be built and used with much less need for retuning. A loop antenna can be mounted on the case in a biocompatible structure. Equations 4 and 5 relate to small and large loops, other equations exist for multi-turn loop designs.

$$R_{rad} = 31200\left(A/\lambda^2\right)^2 \qquad A \leq \lambda^2/100 \tag{4}$$

where $R_{rad}$ is radiation resistance and A is the loop area and λ the wavelength in medium.

$$R_{rad} = 3270\left(A/\lambda^2\right)^2 \qquad A > \lambda^2/100 \tag{5}$$

More details of antenna design can be found from Kraus [16] Fujimoto [17], Lee [18], and Krall [19]. The performance of an implanted communication system within a body is difficult to predict or simulate. Approximation to a human body can be made with a body phantom liquid as described in the book edited by Yang [15]. Unlike applications in air, there are no reliable equations to use and therefore only limited simulation models. That said, simulation can provide a guide to the propagation from a body but should not be used to guarantee performance.

## 3 On-Body Communication

The rapid growth in intelligent sensors, microelectronics and integrated circuit, system-on-chip design, and low power wireless communication has introduced the development of miniaturized and non-invasive sensor nodes. These non-invasive sensor nodes can be placed on human body to create an on-body communication network, which can be used for ambulatory health monitoring of a patient. Unlike in-body communication where the devices are implanted in human body, in on-body communication network, the tiny sensors are placed on the body with out implantation, which provides long term health monitoring and prevents the occurrence of life threatening events. The information is gathered into a central intelligent node or PDA, which also provides an interface to the patient as well as communicates with a remote server. A BSN usually consists of three levels [20]. The first level is called sensor level, which consists of miniaturised low power sensors such as ECG (electrocardiogram), SpO2 (oxygen saturation sensor), EMG (electromyography) and EEG (electroencephalography). The second level called PDA or central intelligent node collects patient information and communicates with the remote station. The third level consists of a remote base station, which keeps patient medical records and provides diagnostic recommendations [20].The GPRS system is used to track the patient's location. A number of on-going projects such as CodeBlue [21], MobiHealth [22] and Connect [23] have facilitated research in on-body communication networks. A system architecture of wireless body area network is presented in [20], where existing Telos platform with an integrated wireless ZigBee compliant radio with on-board antenna is modified by adding ISMP (Intelligent signal processing module) component. This architecture performs real time analysis of sensors data, provides feedback to the user and forwards the user's information to a telemedicine server. A project called Ubiquitous Monitoring Environment for Wearable and Implantable Sensors (UbiMon) aims to develop a smart and affordable health care system and is designed by using six components: the sensors, the remote sensing units, the local processing units, the central server, the patient database, and the workstation [24]. A BSN node for on-body network is developed during this project. The BSN node provides a versatile environment for pervasive healthcare applications and requires 0.01mA in active mode.The BSN node uses IEEE 802.15.4 (Zigbee) wireless link as a low power communication protocol. However, the narrowband implementation doesn't satisfy the energy consumption budget of the sensor nodes and hence, an alternative solution is required. The emerging UWB technology is considered to be the best alternative solution, which could reduce the baseline power consumption of sensor nodes. A pulse-based UWB scheme for on-body communication networks [25],UWB channel measurement with antennas placed on human body [26] and UWB antennas for a BSN [27] have urged researchers to consider UWB technology for communication within on-body networks.

## 4 Multi-agent Technology for BSN

In case of critical condition, the patient's data should be transferred to a remote server for diagnosis and prescription. This requires the development of smart multi-agent system for healthcare services. In most of the projects such as Mobihealth [22], Politechnico[28] and Tele-medicare [29], the patient's medical information are extracted from PDA and forwarded to a central server in hospital using subsequent multi-agent operations. A multi-agent architecture proposed in [30] uses ontology based mobile agent for real time diagnosis. A multi-agent based healthcare system (MAHS) is presented in [31], which is mainly divided into three areas: BSN, Surrogate System and hospital subsystem. The surrogate system connects BSN and hospital subsystem. This multi-agent system is divided into five main agents: Patient Monitoring Agent (PMA), Gate Agent (GA), Supervisor Agent (SA), Manager Agent (MA) and Doctor Agent (DA). The combined operation of these agents provides patient monitoring, real time feedback to the patient and emergency management. PMA collects data from miniaturised sensors and forwards to surrogate system via SA. GA verifies patient's authentication for his requests. SA controls the surrogate system. MA controls the hospital subsystem. DA provides diagnosis and prescription based on the collected data to PMA. All these multi-agent systems for pervasive health care services require further investigation. The management of the huge amount of patient's data and determining patient's condition based on collected data is a challenging issue and requires advance data mining techniques.

## 5 Open Issues and Challenges

A proactive BSN system requires the resolution of many technical issues and challenges such as biosensor design, power scavenging issue, low power RF data paths, scalability, fault tolerance, low power communication protocol, mobility, interoperability, security and privacy. In on-body communication networks, biocompatibility is the most important issue. The biosensor is often placed on human body and its reliability is relied on the interface between the sensor and tissue or blood [32]. A number of biosensors are developed such as ECG sensor based on Telos platform [33], SpO2 sensor and ECG sensor based on a BSN node [34], DNA sensor [26], 3D accelerometers and gyroscope [23] and piezoresistive shear stress sensor [35]. Another important factor is battery life. The solution of some technical issues such as sensor design, RF design and low power MAC protocol contributes to extend the battery life. Lithium based batteries can operate at 1400-3600J/cc and provide long period of operation i.e. from few months to years [36]. A recently developed Sony product "Bio Battery" which generates electricity from sugar can be a promising candidate to solve the power scavenging issue [37]. IMEC developed a thermal micro power generator, which converts thermal energy to electrical energy [38]. The radio interface is also a major challenge and its power consumption in a BSN must be reduced below the energy scavenging limit (100 micro Watt) [25].

The current sensors nodes are mostly based on RF circuit design. Reducing the power consumption of RF transceiver plays a significant role in increasing the lifetime of a sensor. UWB technology is the best solution to increase the operating period of sensors. However the Power Spectral Density (PSD) must be calibrated inside the Federal Communication Commission (FCC) mask for indoor applications. The tiny biosensors wirelessly transmit the collected information to the central intelligent node. Design of a low power and secure communication protocol for a BSN is the most important issue. HTTP protocol is designed to transfer data to remote base station [39]. Chipcon CC2420 uses IEEE 802.15.4 (ZigBee) wireless link for transmitting physiological data between sensors. A cross layer protocol (MAC/Network layer) called Wireless Autonomous Spanning Tree Protocol (WASP) is presented where a spanning tree is set up autonomously and the network traffic is controlled by broadcasting scheme messages over the tree [40]. An extended version of WASP protocol called Cascading Information Retrieval by Controlling Access with Distributed Slot Assignment (CICADA) is presented, which guarantees low delay and good resilience to mobility [41].

## 6 Conclusions and Future Prospects

A Body Sensor Network (BSN) consists of miniaturised, invasive and non-invasive, low power autonomous sensor nodes, which are seamlessly placed on or implanted in human body in order to provide an adaptable and smart healthcare system. A successful BSN system requires the resolution of many technical issues and challenges, which includes but not limited to interoperability, QoS, privacy and security, low power RF data paths, power scavenging issue, biosensor design, scalability and mobility. Moreover, in implant communication, the implant transceiver needs to be sensitive on receive with the ability to tune the antenna for best response. In this paper, we briefly discussed In-body and On-body communication networks. We talked about the methodologies of wireless communication between implanted medical devices with external monitoring equipment. Moreover, we presented a comprehensive discussion on on-body communication networks with a special focus on the recent technological trend in this area. Technical issues and challenges in BSN have also been discussed. Future applications include smart health care services, remote diagnostic and telemedicine, wearable technology to monitor vital signs, smart nursing homes, emergency communication and patient's data maintenance. The broadband signaling scheme such as UWB is a promising candidate to satisfy power consumption budget of sensor nodes and is under investigation in different research institutes. To enable uplink communication from sink to nodes, the WASP and CICADA need to be improved.